\documentclass[letterpaper, 10 pt, conference]{ieeeconf}  

\IEEEoverridecommandlockouts                              

\usepackage[utf8]{inputenc}
\usepackage{xcolor}
\usepackage{amsmath,amssymb,amsfonts}
\usepackage{graphicx}
\usepackage{comment}
\usepackage[short]{optidef}
\usepackage{algorithm}
\usepackage{algpseudocode}
\usepackage{svg}
\usepackage{subfiles}
\usepackage{cite}[sort,compress]

\newcommand{\stox}{{x}}

\newcommand{\stou}{{u}}

\newcommand{\stoz}{{z}}
\newcommand{\tx}[1]{{#1}}

\graphicspath{{\subfix{../images/}}}

\title{\LARGE \bf
Computationally efficient predictive control\\ based on  ANN state-space models
}

\author{Jan H. Hoekstra, Bence Cseppent\H{o}, Gerben I. Beintema, Maarten Schoukens, Zsolt Kollár, and Roland Tóth
\thanks{B. Cseppent\H{o} and Zs. Kollár are with the Dept.~of Measurement and Information Systems at the Budapest University of Technology and Economics, Hungary, while the rest of the authors are with the Control Systems group in the Dept.~of Electrical Engineering at the Eindhoven University of Technology, The Netherlands. Roland T\'oth is also affiliated with the Institute for Computer Science and Control, Budapest, Hungary.}%
    \thanks{Corresponding author: B. Cseppent\H{o} (\texttt{cseppento@mit.bme.hu})}}

\begin{document}

\maketitle
\thispagestyle{empty}
\pagestyle{empty}

\begin{abstract}

Artificial neural networks (ANN) have been shown to be flexible and effective function estimators for identification of nonlinear state-space models. However, if the resulting models are used directly for nonlinear model predictive control (NMPC), the resulting nonlinear optimization problem is often overly complex due the size of the network, requires the use of high-order observers to track the states of the ANN model, and the overall control scheme exploits little of the structural properties or available autograd tools for these models. In this paper, we propose an efficient approach to auto-convert ANN state-space models to linear parameter-varying (LPV) form and solve predictive control problems by successive solutions of linear model predictive problems, corresponding to quadratic programs (QPs). Furthermore, we show how existing ANN identification methods, such as the SUBNET method that uses a state encoder, can provide efficient implementation of MPCs. The performance of the proposed approach is demonstrated via a simulation study on an unbalanced disc system.        \vspace{-1mm}


\end{abstract}

\section{Introduction}

Over the past decades, 
intensive research has been conducted on \emph{model predictive control} (MPC) due the advantages of predictive approaches over reactive feedback solutions and reliable constraint handling, making MPC a core technology in many industrial sectors \cite{rawlings_mpc}. However, current technological developments and increasing performance expectations pushing system designs towards exhibiting more dominant nonlinear behavior, giving rise to increasing need for reliable MPC solutions for \emph{nonlinear} (NL) systems.

Despite of an increasing range of highly competitive NMPC solutions, e.g, \cite{Verschueren2022, Limon2008}, a serious obstacle that one encounters in their application in practice is the required existence of an accurate model of the underlying system. As obtaining models of increasingly complex system designs is becoming infeasible 
via previously-used first principles-based methods, data-driven techniques started to gain serious importance in practice.    To accomplish accurate data-driven modelling of complex NL systems, \emph{machine learning} methods have recently appeared to provide a reliable approach. Especially \emph{artificial neural networks} (ANN) have been shown to be flexible in capturing complicated nonlinear dynamic relationships and effective to deal with the high-complexity of the involved estimation problem. For control applications, ANN \emph{state-space} (SS) models \cite{suykens1995} are particularly attractive and with the recent introduction of encoder-based statistically efficient methods such as \cite{Beintema2022,Forgione2021TruncSimulation}, estimation of them can be reliably and computationally efficiently accomplished in practice. 


While NMPC methods have been applied to various ANN models, e.g., in \cite{lazar2002, Lawrynczuk2014}, these approaches are mainly based on the application of existing general NMPC techniques and linearization based methods. For example, it is not complicated to replace the nonlinear system model in any NMPC scheme with an ANN-SS model and solve the resulting problem with general solvers, e.g., \cite{ipopt2006}. However, ANNs-based models have many nonlinear activation functions which are all needed to be evaluated and linearized by the solver without being aware of the underlying structure, often resulting in excessive computational complexity. Furthermore, the use of high-order observers is also required to track the states of the ANN model, increasing further the computational load. 

In this paper, we propose a computationally efficient iterative MPC scheme for ANN-SS models through the use of automated \emph{linear parameter varying} (LPV) conversion based on the \emph{Fundamental Theorem of Calculus} (FTC) and iterative \emph{quadratic programming} (QP). The computational complexity of this conversion is minimized through a computational structure that exploits the use of highly efficient autograd methods and multi-threading-based computing. Also, the encoder obtained with the identification of the ANN-SS model is used as a direct observer for state estimation. Our contributions can be summarized as follows:
\begin{itemize}
\item Approximation-free auto conversion of ANN-SS models to an LPV form through the FTC, efficiently computed via autograd methods and multi-threaded computing.
\item Direct formulation of successive QP problems using the obtained optimal solution from the previous iteration step as the scheduling for the LPV prediction model.
\end{itemize}
The remainder of the paper is structured as follows. ANN-SS models are introduced in Section~\ref{sec:Model}, followed by a discussion of the corresponding NMPC problem in Section~\ref{sec:NMPC} and the  proposed LPV conversion scheme in Section~\ref{sec:LpvConversion}. The use of the LPV conversion to solve the NMPC efficiently via an iterative scheme is covered in Section~\ref{sec:ANN:MPC}. A simulation example based on an unbalanced disk system is used to demonstrate performance of the proposed MPC approach and compare it to a standard NMPC solution in Section~\ref{sec:Example}.

\vspace{-1mm}
\section{Deep-learning based state-space models}\label{sec:Model}




Consider a \emph{discrete-time} (DT) nonlinear system  in the form: \vspace{-4mm}
\begin{subequations}\label{eq:nl_dyn}
	\begin{align}
		x_{k+1}&=f(x_k,u_k),\\
		y_k&=h(x_k) +e_k,
	\end{align} 
\end{subequations} 
where  $x_k \in \mathbb{R}^{n_\mathrm{x}}$ is the state, $u_k \in \mathbb{R}^{n_\mathrm{u}}$ is the input, $y_k \in \mathbb{R}^{n_\mathrm{y}}$ is the output signal of the system at time moment $k\in\mathbb{Z}$ with $e_k$ an i.i.d. white noise process representing measurement noise, and the state-transition function $f: \mathbb{R}^{n_\mathrm{x}} \times \mathbb{R}^{n_\mathrm{u}} \rightarrow \mathbb{R}^{n_\mathrm{x}}$ and output function $h: \mathbb{R}^{n_\mathrm{x}}\rightarrow \mathbb{R}^{n_\mathrm{y}}$ are bounded functions. 

Using a data sequence  $\mathcal{D}_N=\left\{\left(y_k, u_k\right)\right\}_{k=1}^N$ generated by \eqref{eq:nl_dyn}, we aim to identify a DT model of the form \vspace{-0.5mm}
\begin{subequations}\label{eq:model_structure}
    \begin{align}
        \hat{x}_{k+1} &=f_\theta\left(\hat{x}_k, u_k\right), \\
        \hat{y}_k &=h_\theta\left(\hat{x}_k\right),
    \end{align}
\end{subequations} \vskip -0.5mm \noindent
where $\hat{x}_ k\in \mathbb{R}^{n_{\hat{x}}}$ is the model state, $\hat{y}_k \in \mathbb{R}^{n_\mathrm{y}}$ is the model output, and $f_\theta : \mathbb{R}^{n_{\hat{\mathrm{x}}}} \times \mathbb{R}^{n_\mathrm{u}} \rightarrow \mathbb{R}^{n_{\hat{\mathrm{x}}}}$ with $h_\theta: \mathbb{R}^{n_{\hat{\mathrm{x}}}}  \rightarrow \mathbb{R}^{n_{\mathrm{y}}}$ are feedforward multi-layer artificial neural networks\footnote{Consider that each hidden layer is composed from $m$ activation functions $\phi:\mathbb{R} \rightarrow \mathbb{R}$ in the form of $\stoz_{i,j} = \phi(\sum_{l=1}^{m_{i-1}}\theta_{\mathrm{w},i,j,l} \stoz_{i-1,l}+ \theta_{\mathrm{b},i,j})$ where $\stoz_i=\mathrm{col}(z_{i,1},\ldots,z_{i,{m_i}})$  is the latent variable representing the output of layer $1\leq i\leq q$. Here, $\mathrm{col}(\centerdot)$ denotes the composition of a column vector. For $f_\tx{\theta}$ with $q$ hidden-layers and linear input and output layers, this means $f_\tx{\theta}(\hat{\stox}_\tx{k}, \stou_\tx{k})= \theta_{\mathrm{w},q+1} \stoz_q(k) + \theta_{\mathrm{b},q+1}$ and $\stoz_{0}(k)=\mathrm{col}(\hat{x}_\tx{k}, \stou_\tx{k})$. } with $\theta\in\mathbb{R}^{n_\theta}$ collecting the activation weights as model parameters.  Under these considerations, 
\eqref{eq:model_structure} represents a recurrent neural network, also called as an ANN-SS model \cite{suykens1995}. 

\begin{figure}[t]
    \centering
    \includegraphics[width=0.95\linewidth]{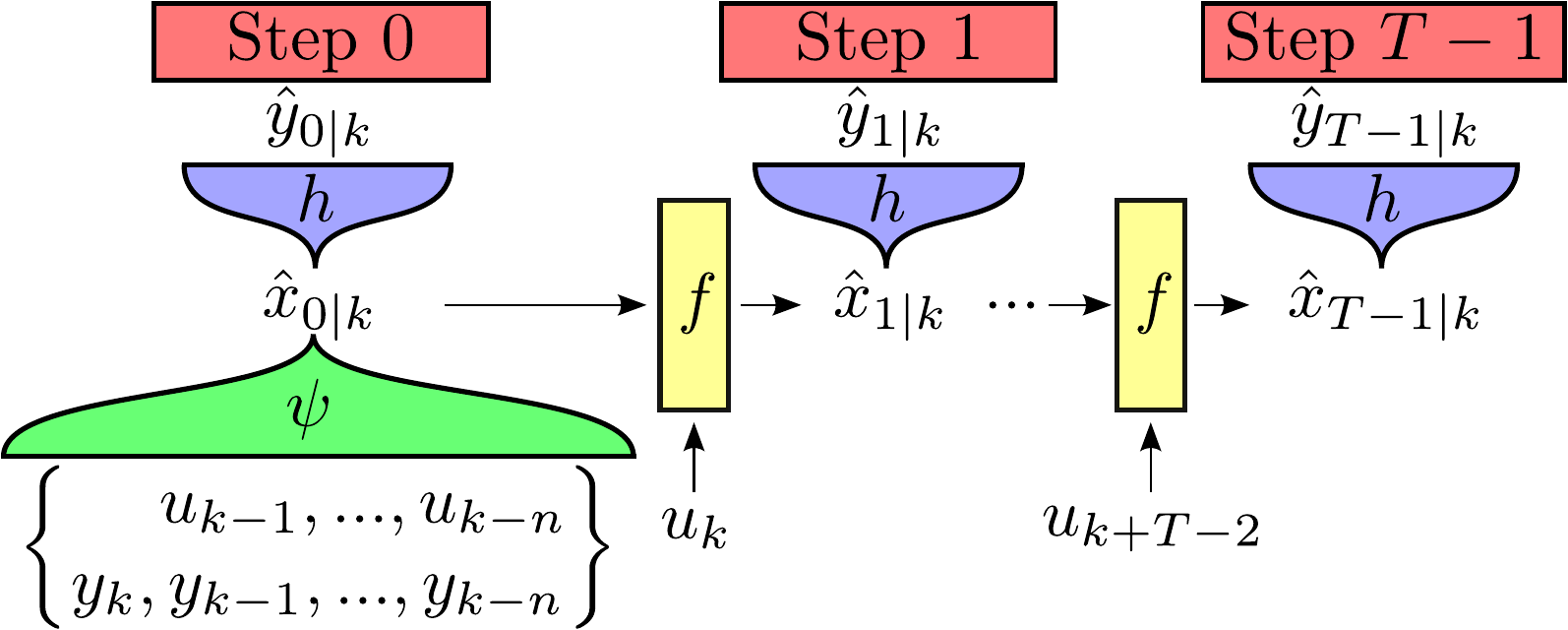}
    \caption{SUBNET structure: the subspace encoder $\psi_\theta$ estimates the initial state at time index $k$ based on past inputs and outputs, then the state is propagated through  $f_\theta$ and $h_\theta$ multiple times until a simulation length $T$.}
    \label{fig:n-step-encoder-graphic} \vspace{-6mm}
\end{figure}

In the literature, many approaches have been introduced to identify ANN-SS models such as \cite{masti2021autoencoders,Forgione2021TruncSimulation}. One recent approach that provides computationally efficient estimation of \eqref{eq:model_structure} under statistical consistency guarantees is the SUBNET approach \cite{Beintema2022}. To achieve computationally efficient estimation, SUBNET uses a truncated prediction cost as the objective function to be minimized during identification which also enables the use of batch optimisation methods. This truncation corresponds to a forward simulation of \eqref{eq:model_structure} over multiple subsections of length $T$ of the available data 
as depicted in Figure \ref{fig:n-step-encoder-graphic}, starting from randomly selected time instances $k\in \{n+1,\ldots,N-T\}$. In order to compute such simulation batches, SUBNET co-estimates an ``encoder'' function $\psi_\theta$ together with $f_\theta$ and $h_\theta$, which provides a direct estimate of the initial state $\hat{x}_{0|k}$ based on past input and output data, i.e., $\hat{x}_{0|k}=\psi_\theta(u_{k-n}^{k-1},y_{k-n}^{k})$ where $u_{k}^{k+\tau}= [\ u_k^\top \ \cdots \ u_{k+\tau}^\top ]^\top$ for $\tau\geq 0$ and $y_{k}^{k+\tau}$ is defined similarly. 
Next, we formulate the predictive control problem we intend to solve for identified ANN-SS models in the form of \eqref{eq:model_structure}.   

\vspace{-1mm}  
\section{Nonlinear predictive control problem} \label{sec:NMPC}





Traditional NMPC corresponds to repeatedly solving an \emph{optimal control problem} (OCP) on a finite prediction horizon 
using the current observation of the state variables $x_k$ as an initial value, then applying the first element of the acquired input sequence to the system, 
see~\cite{rawlings_mpc}. 
However, we consider the true NL system equations~\eqref{eq:nl_dyn} to be unknown and to have only an approximate ANN-SS model \eqref{eq:model_structure} with estimated parameters $\hat\theta$. Considering a regulation objective w.r.t. a set point $x_\mathrm{ref}\in\mathbb{R}^{n_\mathrm{x}}$, NMPC problem based on \eqref{eq:model_structure}  and under a quadratic cost is formulated as follows: \vspace{-1mm}
\begin{subequations}
\label{eq:nl_ocp}
\begin{align}
    \min_{{u}_{0|k}^{N_\mathrm{p}-1}} \ & \sum_{i = 1}^{N_\mathrm{p}} \ell_i(\hat{x}_{i|k}\!-\!x_{\textrm{ref}},{u}_{i-1|k}\!-\!u_{\textrm{ref}}),  
    \label{eq:NLcost}\\
    \textrm{s.t.} \ \ & \hat x_{i+1|k} = f_{\hat{\theta}}(\hat x_{i|k},u_{i|k}), \quad i \in\mathbb{I}_0^{N_\mathrm{p}-1}, \\
    & \hat y_{i|k} = h_{\hat{\theta}}(\hat x_{i|k}), \hspace{14mm} i \in\mathbb{I}_1^{N_\mathrm{p}}, \\
      &  u_{\textrm{min}} \leq {u}_{i|k}\leq {u}_{\textrm{max}}, \hspace{8.5mm} i \in\mathbb{I}_0^{N_\mathrm{p}-1}, \label{eq:u_const} \\
      &y_{\textrm{min}} \leq \hat {y}_{i|k} \leq y_{\textrm{max}},  \hspace{9.5mm} i \in\mathbb{I}_1^{N_\mathrm{p}} ,  \label{eq:y_const} \\
      &\hat x_{0|k} = \hat  x_k, \label{eq:inital}
\end{align}
\end{subequations}\vskip -1mm \noindent 
\noindent where subscript $(i|k)$ means the predicted value of the variable at $i+k$ in the optimisation problem when the current time is $k\in\mathbb{Z}$, $\ell_i(x,u)=x^\top\! Q_i x + u^\top\! R_i u$ is the quadratic stage cost where $Q_i$ are positive definite and $R_i$ are positive semi-definite weighting matrices defining the user-chosen performance specification for the controller, $\mathbb{I}_{\tau_1}^{\tau_2}=\{k\in\mathbb{Z} \mid \tau_1\leq k \leq \tau_2\}$ is an index set, 
while $y_{\mathrm{min}},\ldots,u_{\mathrm{max}}$ correspond to box constraints\footnote{The use of box constraints is an arbitrary choice, in fact, any polyhedral set can be considered.} on the outputs and the inputs where $\leq$ and $\geq$ is considered element-wise, and $N_\mathrm{p}$ is the length of the prediction horizon (considered to be equivalent with the control horizon). As the model of the true system is unknown and the identified NL-SS model can be on arbitrary state basis, constraints imposed on the states would be meaningless. Therefore, only output constraints in \eqref{eq:y_const} are considered. On the other hand, to realize the set-point objective, it is assumed that a set point target $(x_\mathrm{ref},u_\mathrm{ref})$ is synthesised for example based on the approach in \cite{GonzalezCisneros2021}. Considering the tracking objective in terms of $x_\mathrm{ref}$ also makes possible to ensure stability of the predictive controller for example via terminal ingredients \cite{GonzalezCisneros2021}, however, due to the lack of space, this is not considered in the paper.

Even under these considerations, the main challenge faced by existing NMPC methods applied on \eqref{eq:nl_ocp} is that fast online solutions of this nonlinear problem are required under potentially large state-dimensions and heavily complex functions $f_{\hat \theta}$ and $h_{\hat \theta}$, which are required to be passed analytically to the existing NMPC solvers. Hence, it becomes attractive to develop a dedicated predictive control solution that can exploit the structural
properties of ANN-SS models and \texttt{autograd} tools to provide accelerated solutions for \eqref{eq:nl_ocp}. 
\vspace{-4mm}
\section{Conversion to surrogate LPV form} \label{sec:LpvConversion}
To provide an efficient solution of \eqref{eq:nl_ocp}, we aim to convert the ANN-SS model \eqref{eq:model_structure} with estimated parameters $\hat \theta$ to an LPV form. We will show in Section \ref{sec:ANN:MPC} that this form enables to solve \eqref{eq:nl_ocp} via a chain of computationally efficient QPs.   
To realize this objective, we need an efficient auto-conversion of an ANN-SS model \eqref{eq:model_structure} to an LPV-SS form \vspace{-4mm}
\begin{subequations}
\label{eq:lpv_dyn}\begin{align}
        \hat x_{k+1} &= A(p_k) \hat x_k + B(p_k)  u_k,\\
        y_k &= C(p_k) \hat x_k,  
\end{align}
\end{subequations} 
where $p_k$ is the so-called \emph{scheduling variable}, dependent on $\hat{x}_k$ and $u_k$ in terms of the \emph{scheduling map} $p_k=\mu(\hat{x}_k,u_k)$. Note that \eqref{eq:lpv_dyn} corresponds to $f_{\hat\theta}(\hat{x},u)=A(\mu(\hat{x},u))\hat x+B(\mu(\hat{x},u))u$ and $h_{\hat\theta}(\hat{x},u)=C(\mu(\hat{x},u))\hat x$, 
which are non-unique factorizations of the nonlinearities. The map $\mu$ is often synthesized under the expectation that $A,\ldots,C$ belong to a given function class, e.g. affine, polynomial, rational, etc.

In the literature, a wide range of LPV conversion methods for general NL-SS models is available, however, many of these methods are either based on complex analytical formulas and restricted to narrow system classes, e.g., \cite{hashemi2012,Marcos04} or follow a data-driven conversion approach, e.g. \cite{Kwiatkowski4494453,Toth20bIET}, which makes them difficult to automate and inefficient for conversion of potentially large-scale ANN models. 
A novel method that has been recently derived in~\cite{Koelewijn2023} is based on the \emph{Fundamental Theorem of Calculus} (FTC) and, as we show here, can be used to exploit \texttt{autograd} tools and parallel computation to recast general ANN-SS models as~\eqref{eq:lpv_dyn}. 


The core idea is that if given a continuously differentiable function $g:\mathbb{R}^n\rightarrow\mathbb{R}^m$, then based on the FTC~\cite{Koelewijn2023}:
\begin{align}
  g(\eta)- g(0) &=\left(\int_0^1\frac{d g}{d\eta}({\lambda}\eta)\,d{\lambda}\right)\eta,
\end{align}
where $\frac{d g}{d\eta}({\lambda}\eta) \in \mathbb{R}^{m \times n}$ is the Jacobian of $g$ evaluated in ${\lambda}\eta$. Provided that functions $f$ and $h$ in \eqref{eq:nl_dyn} are differentiable, choosing  $\eta=[\
	\hat{x}^\top \ \ u^\top \ ]^\top$ gives via the FTC that
    \begin{align}
	\tilde{f}_{\hat \theta}(\hat{x},u)&\!=\!\underbrace{  \left({\int_0^1\!\!\frac{\partial f}{\partial \hat{x}}({\lambda}\hat{x},{\lambda}u)\,d\lambda}\right)}_{{A}(\hat{x},u)}\!\hat{x}\!+\!\underbrace{\left(\int_0^1\!\!\frac{\partial f}{\partial u}({\lambda}\hat{x},{\lambda}u)\,d\lambda\right)}_{{B}(\hat{x},u)}\!u, \notag \\
	\tilde{h}_{\hat \theta}(\hat{x})&\!=\!\underbrace{\left(\int_0^1\!\!\frac{\partial h}{\partial \hat{x}}({\lambda}\hat{x})\,d\lambda\right)}_{{C}(\hat{x})}\!\hat{x}, \label{eq:abcd_def} 
   \end{align}
\vskip -2mm \noindent where $	\tilde{f}_{\hat \theta}(\hat{x},u)=f_{\hat \theta}(\hat{x},u)-	f_{\hat \theta}(0,0)$ with $\tilde{h}_{\hat \theta}$ similarly defined. With $p_k=[\
	\hat{x}_k^\top \ \ u_k^\top \ ]^\top$ and $\mu$ the identity function, this gives an LPV form of the ANN-SS model \eqref{eq:model_structure} with affine terms $V=	f_{\hat \theta}(0,0)$ and  $W=	h_{\hat \theta}(0)$ as
	\begin{subequations}
\label{eq:lpv_model}\begin{align}
      \hat{x}_{k+1} &= A(p_k) \hat{x}_k + B(p_k)  u_k + V,\\
        y_k &= C(p_k)\hat{x}_k 
        + W. 
\end{align}
\end{subequations} 
While calculating the Jacobians and solving the integrals to obtain $A,\dots,C$ is a difficult task analytically, we will show that for an MPC implementation, these matrix functions are needed to be computed only for specific values of $x$ and $u$, hence algorithmic differentiation and a suitable numerical integration scheme (e.g., Simpson rule) can be efficiently used for this purpose. 
Furthermore, for ANNs, where the activation functions are not everywhere differentiable over their argument space, but they are Lipschitz continuous, \texttt{autograd} methods still provide a subgradient, see \cite{shalev2014Subgradient}. Next, we  show how this LPV surrogate form of the identified ANN-SS model can be used to solve \eqref{eq:nl_ocp} efficiently.

\section{Iterative ANN-MPC method}\label{sec:ANN:MPC}

In this section, we develop an iterative optimisation algorithm to solve the ANN-SS model based predictive problem \eqref{eq:nl_ocp} based on the concept of the so called "quasi" LPV-MPC approach \cite{GonzalezCisneros2021}. The core idea is that, at any given time moment $k$, for a fixed scheduling sequence $\{p_{i+k}\}_{i=0}^{N_\mathrm{p}}$, \eqref{eq:lpv_model} is used to formulate a linear MPC problem that can be solved rapidly as a QP. Then the resulting  control sequence $\{{u}_{i+k}\}_{i=0}^{N_\mathrm{p}-1}$ is used to forward simulate \eqref{eq:nl_dyn} to compute a new sequence $p_{i+k}=[\ \hat{x}_{i+k}^\top \ \  {u}_{i+k}^\top\ ]^\top$ on which a new linear MPC problem is formulated and solved. Execution of this iteration multiple times until a converging $u_{i+k}$ trajectory is reached   ensures that the LPV representation sufficiently well approximates the NL system along a (locally) optimal solution trajectory of \eqref{eq:nl_ocp} for which the solution of the linear MPC problem coincides with the solution of \eqref{eq:nl_ocp}.      


\subsection{Reformulation to a quadratic problem}\label{sec:ReformulationToQP}

To exploit the above sketched idea, the NMPC problem \eqref{eq:nl_ocp} at time moment $k\in \mathbb{Z}$, for a given fixed scheduling sequence $\{p_{i|k}\}_{i=0}^{N_\mathrm{p}}$  is reformulated as the LPV-MPC problem
\begin{subequations}
\label{eq:LPV_ocp}
\begin{align}
\hspace{-2mm}    \min_{\breve{u}_{0|k}^{N_\mathrm{p}-1}} \ & \sum_{i = 1}^{N_\mathrm{p}} \ell_i(\breve{x}_{i|k}\!-\!x_{\textrm{ref}},\breve{u}_{i-1|k}\!-\!u_{\textrm{ref}}), \label{eq:LPVcost}\\
    \textrm{s.t.} \ \ &  \breve x_{i+1|k} \!=\! A(p_{i|k}) \breve{x}_{i|k}\!+\! B(p_{i|k})  \breve{u}_{i|k}\! +\! V,\hspace{0.7mm} i\!\in\!\mathbb{I}_0^{N_\mathrm{p}-1}  
    \label{eq:LPV:state:prop} \\
    & \breve y_{i|k}= C(p_{i|k})\breve{x}_{i|k}+ W, \hspace{7mm} i \in\mathbb{I}_1^{N_\mathrm{p}},  \label{eq:LPV:output:prop} \\
      &  u_{\textrm{min}} \leq \breve{u}_{i|k}\leq {u}_{\textrm{max}}, \hspace{13.5mm} i \in\mathbb{I}_0^{N_\mathrm{p}-1}, \label{eq:LPV:u_const} \\
      &y_{\textrm{min}} \leq  \breve{y}_{i|k} \leq y_{\textrm{max}},  \hspace{14.5mm} i \in\mathbb{I}_1^{N_\mathrm{p}} ,  \label{eq:LPV:y_const} \\
      &\breve x_{0|k} =x_k, \label{eq:LPV:inital}
\end{align}
\end{subequations}
Using linearity of \eqref{eq:LPV:state:prop} and \eqref{eq:LPV:output:prop} along $\{p_{i|k}\}_{i=0}^{N_\mathrm{p}}$ gives 
\begin{subequations} \label{eq:lpv:pred:Model}
\begin{align} 
\breve X_{k+1}&=\Phi\left({P}_k\right) \breve x_{0|k} +\Gamma\left(P_k\right) \breve U_k + {F}_0({P}_k) V,\\
\breve Y_{k+1}&={\Lambda}\left({P}_{k+1}\right) \breve X_{k+1} + {H}, \label{eq:lpv:pred:Model:y}
\end{align}
\end{subequations}
where $\breve X_{k+1}= \mathrm{col}(\{\breve{x}_{i|k}\}_{i=1}^{N_\mathrm{p}})= [\begin{array}{ccc} \breve{x}_{1|k}^\top & \cdots & \breve{x}_{N_\mathrm{p}|k}^\top \end{array}]^\top $ and $\breve U_k = \mathrm{col}(\{\breve{u}_{i|k}\}_{i=0}^{N_\mathrm{p}-1})$ with $\breve Y_{k+1}$ and $P_k$ similarly defined. $\Phi\left(\mathrm{P}_k\right)$ and $\Gamma\left(\mathrm{P}_k\right)$ contain various products of  $A(p_{i|k})$ and  $B(p_{i|k})$ with a structure as discussed in \cite{GonzalezCisneros2021},\vspace{-2mm}\begin{equation*}
  {F}_0({P}_k)= I_{n_\mathrm{x}}  + \left[\begin{array}{cccc}
    0 &\hspace{-1mm} A^\top\!(p_{1|k}) & \sum_{\tau=1}^{N_\mathrm{p}-1} \prod_{i=1}^{\tau} A^\top\!(p_{i|k})
    \end{array}\right]^\top
\end{equation*}
with $I_{n_\mathrm{x}}$ corresponding to an identity matrix of dimension ${n_\mathrm{x}}$. Furthermore, ${\Lambda}\left({P}_{k+1}\right) =\mathrm{diag}(C(p_{1|k}),\ldots, C(p_{N_\mathrm{p}|k}))$ and $H=\mathrm{col}(W,\ldots,W)$.
Based on \eqref{eq:lpv:pred:Model}, the objective function \eqref{eq:LPVcost} can be written as
\begin{equation} \label{eq:CostNonConstrained}
    {J}_k = \breve U_k^\top G\left({P}_k\right) \breve U_k + F^\top\!\! \left({P}_k\right) \breve U_k,
\end{equation}
similar to  \cite{GonzalezCisneros2021}, where $G\left({P}_k\right)\! =\! 2\left(\Psi \!+\! \Gamma^\top  \!\! \left({P}_k\right)\Omega \Gamma\left({P}_k\right) \right)$ and 
\begin{multline}
    F({P}_k)  = 2\left(\Gamma\left({P}_k)^\top \Omega (\Phi({P}_k) \hat{x}_{0|k}  \right. \right.+\\
     + {F}_0({P}_k) V  \left. \left.
    - X_\mathrm{ref}\right) - \Psi^\top U_\mathrm{ref}\right)
    \end{multline}
with $\Omega=\mathrm{diag}(Q_1,\ldots,Q_{N_\mathrm{p}})$, $\Psi=\mathrm{diag}(R_1,\ldots,R_{N_\mathrm{p}})$,  $X_\mathrm{ref}=\mathrm{col}(x_\mathrm{ref},\ldots,x_\mathrm{ref})$ and $U_\mathrm{ref}=\mathrm{col}(u_\mathrm{ref},\ldots,u_\mathrm{ref})$. This gives that, along a fixed trajectory of $\{p_{i|k}\}_{i=0}^{N_\mathrm{p}}$, the unconstrained  part of the predictive problem for the ANN-SS model can be efficiently recasted as a single QP \eqref{eq:CostNonConstrained} in the decision variables $\breve U_k$. 

\vspace{-1mm}
\subsection{Reformulation of the constraints} \label{sec:constraints}


To add the constraints \eqref{eq:LPV:u_const} and \eqref{eq:LPV:y_const} to the QP form \eqref{eq:CostNonConstrained} of the optimization problem, they need to be rewritten in the variable $\breve U_k$. We start by writing the constraints as
\begin{equation}
    M_i \breve{y}_{i+1|k} + E_i \breve u_{i|k} \leq b_i,\quad  \forall i \in \mathbb{I}_0^{ N_\mathrm{p}} \label{eq:const:step} \vspace{-2mm}
\end{equation}
where \vspace{-2mm}
$$ M_i\! =\!
    \begin{bmatrix}
    0_{2n_\mathrm{u} \times n_\mathrm{y}} \\
    -I_{n_\mathrm{y}} \\
    I_{n_\mathrm{y}}
    \end{bmatrix},\  E_i\! =\!
    \begin{bmatrix}
    -I_{n_\mathrm{u}} \\
    I_{n_\mathrm{u}} \\
    0_{2 n_\mathrm{y}  \times n_\mathrm{u}}
    \end{bmatrix},\ b_i\! =\!
    \begin{bmatrix}
    -u_\mathrm{min} \\
    u_\mathrm{max} \\
    -\hat{y}_\mathrm{min} \\
    \hat{y}_\mathrm{max}
    \end{bmatrix}. \vspace{-1mm}
    $$
   Putting  \eqref{eq:const:step} together gives 
\begin{equation} \label{eq:constraints}
\mathcal{M} \breve Y_{k+1}+\mathcal{E} \breve U_k \leq c ,
\end{equation}
where the matrices are given as  $\mathcal{M} =\mathrm{diag}(M_1,\ldots,M_{N_\mathrm{p}})$, $\mathcal{E} =\mathrm{diag}(E_1,\ldots,E_{N_\mathrm{p}})$,  and $c=\mathrm{col}(b_1,\ldots,b_{N_\mathrm{p}})$. Substituting \eqref{eq:lpv:pred:Model} into \eqref{eq:constraints} gives 
\begin{equation} \label{eq:const:final}
L\left({P}_{k}, {P}_{k+1}\right) \breve U_k \leq c+W\left({P}_{k}, {P}_{k+1}\right),
\end{equation}
which is a linear constraint in $\breve U_k$ with matrices
\begin{subequations}
\begin{align}
L\left({P}_{k},{P}_{k+1}\right) =&\mathcal{M}\Lambda\left({P}_{k+1}\right) \Gamma\left({P}_{k}\right)+\mathcal{E}, \\
W\left({P}_{k},{P}_{k+1}\right) =&
- \mathcal{M} \left(\Lambda\left({P}_{k+1}\right) \Bigl(\Phi\left({P}_{k}\right) \hat{x}_{0|k} \right. \\
& \left. +{F}_0\left({P}_{k}\right) V\Bigl) + H\right).
\end{align}
\end{subequations}
To avoid feasibility 
problems due to model inaccuracies, the constraint can also be softened by a slack variable. 



\vspace{-1mm}
\subsection{Iterative algorithm}
\begin{algorithm}[b] 
\begin{algorithmic}[1]
        \State \textbf{initialization}: Let $k=0$ and  $\{\hat{x}_{i|k} = \hat x_0, \hat{u}_{i|k}=0\}_{i=0}^{N_\mathrm{p}}$
        \While{$k \leq N_\text{sim}$}
        \Repeat
        \State \textbf{update} $p_{i|k}$ by $\mu(\hat{x}_{i|k} , \hat{u}_{i|k})$ for $i\in\mathbb{I}_{i=0}^{N_\mathrm{p}}$
        \State \textbf{calculate} $A(p_{i|k}),\ldots,C(p_{i|k})$ via \eqref{eq:abcd_def} 
        \State \textbf{solve} \eqref{eq:LPV_ocp} to obtain $\{\breve{u}_{i|k}\}_{i=0}^{N_\mathrm{p}-1}$
        \State \textbf{simulate} \eqref{eq:model_structure} with $\hat u_{i|k}\! \leftarrow\! \breve{u}_{i|k}$ to obtain $\{\hat{x}_{i|k} \}_{i=1}^{N_\mathrm{p}}$
        \Until $\{\hat{u}_{i|k}\}_{k=0}^{N_\mathrm{p}-1}$ has converged
        \State \textbf{apply} $u_k=\hat{u}_{1|k}$ and observe $\hat x_{k+1}$
        \State \textbf{let} $k\leftarrow k+1$ 
        \State \textbf{propagate} $(\hat{x}_{i|k-1},\hat{u}_{i|k-1})$ to initialize $(\hat{x}_{i|k},\hat{u}_{i|k})$
        \EndWhile 
\caption{Iterated LPV-MPC solution of \eqref{eq:nl_ocp} 
\label{algo:LPV_MPC}}
\end{algorithmic}
\end{algorithm}
As we discussed in Sections \ref{sec:ReformulationToQP} and \ref{sec:constraints}, the original  NMPC problem \eqref{eq:nl_ocp} can be efficiently reformulated as the QP \eqref{eq:CostNonConstrained} with a linear constraint \eqref{eq:const:final} along a given scheduling sequence $\{p_{i|k}\}_{i=0}^{N_\mathrm{p}}$. If this scheduling sequence would be taken as the optimal solution of \eqref{eq:nl_ocp}, i.e., $p_{i|k}=\mu(\hat{x}^\text{NL}_{i|k},\hat{u}^\text{NL}_{i|k})$, the optimal solution of \eqref{eq:LPV_ocp} would coincide with $\{\hat{x}^\text{NL}_{i+1|k},\hat{u}^\text{NL}_{i|k}\}_{k=0}^{N_\mathrm{p}-1}$.
 Hence, by starting from an initial sequence of $\{\hat{x}_{i+1|k},\hat{u}_{i|k}\}_{k=0}^{N_\mathrm{p}-1}$ used to compute $\{p_{i|k}\}_{i=0}^{N_\mathrm{p}}$, the idea is to converge through successive LPV formulations of the NL problem towards the optimal solution of \eqref{eq:nl_ocp} by iteratively solving  \eqref{eq:LPV_ocp} via \eqref{eq:CostNonConstrained} with \eqref{eq:const:final} and use the obtained input sequence $\{\hat{u}_{i|k}\}_{k=0}^{N_\mathrm{p}-1}$ to simulate \eqref{eq:model_structure} and based on the resulting $\{\hat{x}_{i|k}\}_{k=1}^{N_\mathrm{p}}$ update $\{p_{i|k}\}_{i=0}^{N_\mathrm{p}}$. The iteration is continued till a converging input trajectory is achieved and then $u_{0,k}$ is applied to the system in a receding horizon sense. The resulting iterative scheme is summarized in  Algorithm \ref{algo:LPV_MPC}. Iterations often quickly converge in practice and after the computation of the first control cycle at $k=0$, the next cycles can be bootstrapped by taking the optimal trajectory obtained in the previous cycle to initialise $p$, i.e., $\{p_{i|k}=\mu(\breve{x}_{i+1|k-1},\breve{u}_{i+1|k-1})\}_{i=0}^{N_\mathrm{p}-2}$, $p_{N_\mathrm{p}-1|k}=\mu(\breve{x}_{N_\mathrm{p}|k-1},\breve{u}_{N_\mathrm{p}-1|k-1})$, and $p_{N_\mathrm{p}|k}=\mu(\breve{x}_{N_\mathrm{p}|k-1},0)$. Note that $p_{N_\mathrm{p}|k}$ is only used to compute $C$ in \eqref{eq:LPV:output:prop} for \eqref{eq:LPV:y_const} and $C$ is only dependent on the $x$ part of $p$.

\vspace{-1mm}
\subsection{Target selection} \label{target_selector}

In order to realize a set-point tracking objective for the ANN-SS model based MPC problem \eqref{eq:nl_ocp}, $(x_\mathrm{ref},u_\mathrm{ref})$ are required to be determined. In case the objective is defined in terms of a reference $r\in\mathbb{R}^{n_\mathrm{y}}$ that the output needs to follow, the corresponding $(x_\mathrm{ref},u_\mathrm{ref})$ setpoint can be determined via a target selector, for example the method discussed in  \cite{GonzalezCisneros2021}:
\begin{subequations}
\label{eq:target:select}
\begin{equation}
\min _{x_{\textrm{ref}}, u_{\textrm{ref}}} \frac{1}{2}\left\|C\left(p_\textrm{ref}\right) x_{\textrm{ref}}+W-r\right\|_{{Q}}^2+\left\|u_{\textrm{ref}}\right\|_{{R}}^2
\end{equation}
\noindent subject to
\begin{equation}
{\left[\begin{array}{cc}
I-A\left(p_\textrm{ref}\right) & -B\left(p_\textrm{ref}\right) \\
C\left(p_\textrm{ref}\right) & 0
\end{array}\right]\!\left[\begin{array}{l}
x_{\textrm{ref}} \\
u_{\textrm{ref}}
\end{array}\right]=\left[\begin{array}{c}
V \\
\! r-W \!
\end{array}\right]}\vspace{-1mm}
\end{equation}
\begin{equation}
    y_\text{min} \leq C\left(p_\textrm{ref}\right) x_{\textrm{ref}}+W \leq y_\text{max} \vspace{-2mm}
\end{equation}
\begin{equation}
    u_\text{min} \leq u_{\textrm{ref}} \leq u_\text{max},
\end{equation}
\end{subequations}
where \eqref{eq:target:select} can be solved as a QP iteratively by following the procedure used in \ref{sec:ReformulationToQP} and exploiting that $p_\textrm{ref}=\mu(x_\textrm{ref},u_\textrm{ref})$. Due to the efficient QP formulation, the selector can be also executed online in case the reference $r$ varies with $k$.
\vspace{-1mm}
\subsection{Encoder based state observer}





To solve the optimization problem \eqref{eq:LPV_ocp}, the initial state $\hat{x}_{0|k}=\hat{x}_k$ is required. However, from the real system \eqref{eq:nl_dyn}, we can only obtain a noise contaminated output measurement $y_k$. Therefore we require an observer that estimates $\hat{x}_k$ from past values of $u_k$, and the past and current values of $y_k$. For this purpose, various methods can be applied, but if the SS-ANN has been obtained via the SUBNET method, then we can make use the estimated encoder for the model, see Figure \ref{fig:n-step-encoder-graphic}. This encoder is trained under noisy data such that the multiple shooting based prediction loss of the entire model is minimized under noisy measurements of $y_k$. Hence the trained encoder provides filtering and state reconstruction for a reliable estimate of $\hat{x}_k$ and can be used to calculate the initial state as $\breve x_{0|k} =\psi_{\hat \theta}(u_{k-n}^{k-1},y_{k-n}^{k})$ in \eqref{eq:LPV:inital}.

\vspace{-1mm}
\subsection{Computational structure}





\begin{figure}[t]
    \centering
    \includegraphics[width=0.5\textwidth]{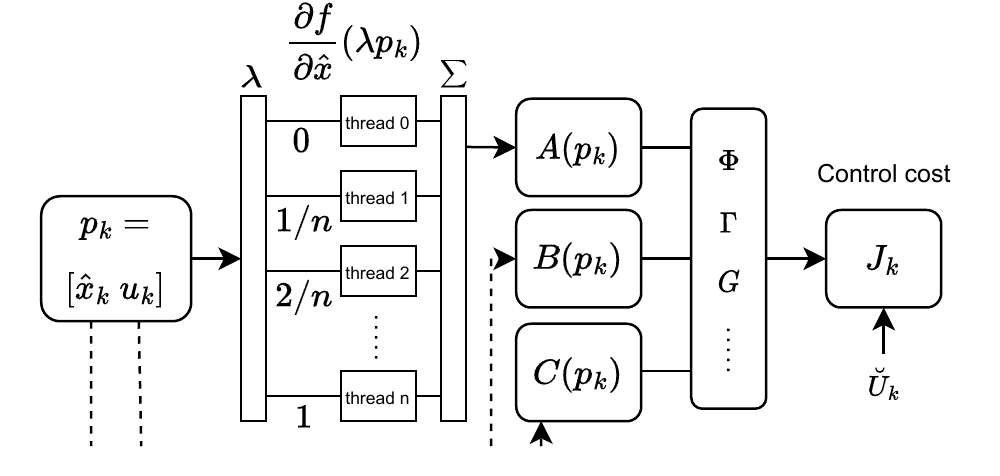} 
    \vspace{-8mm}
    \caption{Rapid calculation of the LPV MPC objective function $J_k$ in \eqref{eq:CostNonConstrained} is executed via autograd calculation of the gradients and numerical evaluation of the integrals in \eqref{eq:abcd_def} using multi-threading. Each thread computes an evaluation of the derivative term for various values of $\lambda$ in the numerical integration scheme and afterwards these results are efficiently summed together to compute each matrix in  $J_k$.} \vspace{-1mm}
    \label{fig:parallelization}
\end{figure}

To set up the QP at each $k$ time moment, the matrices required for \eqref{eq:CostNonConstrained} and \eqref{eq:const:final} are needed to be rapidly computed, which involves computation of $A(p_{i|k}),B(p_{i|k}),C(p_{i|k})$ in terms of \eqref{eq:abcd_def} along a given $\{p_{i|k}\}_{i=0}^{N_\mathrm{p}}$. The involved integrals are numerically computed by a numerical integration scheme (e.g., Simpson rule) corresponding to $n$ evaluation of Jacobians of the ANNs composing the model. Hence, in every iteration in Algorithm \ref{algo:LPV_MPC}, $N_\mathrm{p} \cdot n $ Jacobians are needed to be determined. Due to the lack of dependence between Jacobian calculation, this can be done through \texttt{autograd} methods effectively, and the corresponding $n$ calculations of the Jacobians can be multi-threaded resulting in a highly efficient computational scheme shown in Figure \ref{fig:parallelization}.

\vspace{-1mm}
\subsection{Convergence properties}


In the literature, it has been established that the iteratively solved LPV-MPC scheme is equivalent to  standard Newton-based \emph{sequential quadratic progrmaming} (SQP) if the LPV model \eqref{eq:lpv_model} is obtained using purely Jacobian linearization. This corresponds to the case when, in \eqref{eq:abcd_def}, the integration is dropped and the corresponding $A,B,C$ matrices are approximated only by the Jacobian evaluated at $(\hat{x}_k,u_k)$ \cite{Hespe2021}. Under this simplification, it is sufficient to borrow results from the analysis of Newton SQPs \cite{boggs_tolle_1995}, which are locally convergent if all inequality constraints are predefined. In our algorithm, the latter condition is satisfied by the inequality constraints \eqref{eq:LPV:u_const} and  \eqref{eq:LPV:y_const}. 
Furthermore, for space limitations, the cost function \eqref{eq:LPVcost} does not contain terminal ingredients, but they can be easily incorporated in the problem. In that case, a feasible terminal set would have to be defined to ensure convergence.
Extension of the Newton SQP properties for the exact factorisation provided by \eqref{eq:abcd_def} are expected to hold, while, due to lack of approximation error in \eqref{eq:abcd_def}, this novel variant of SQP is expected to have a faster convergence rate.

\vspace{-1mm}
\section{Example}\label{sec:Example}

Consider a simulator of a nonlinear unbalanced disc system~\cite{Kulcsar2009}. The motion model of the system is 
\begin{subequations}\label{eq:unbalanced}
    \begin{align}
        \dot{\theta}_t &= \omega_t,\\
        \dot{\omega}_t &= \frac{Mgl}{J} \sin(\theta_t) - \frac{1}{\tau} \omega_t + \frac{K_\mathrm{m}}{\tau} v_t,\\
        y_{k} &= \theta_{kT_\mathrm{s}} + e_{k},
    \end{align}
\end{subequations}
where $t\in\mathbb{R}$ is the continous time, $\theta$ is the angle of the disk w.r.t.~the bottom position and measured clockwise, $\omega$ is the angular velocity of the disk, $v$ is the system input in V actuated in a \emph{zero-order hold} (ZOH) sense, $y$ is the DT output measurement with sampling time $T_\mathrm{s}=0.1$ while $e$ is a discrete white noise process with $e_k\sim\mathcal{N}(0,\sigma_\mathrm{e}^2)$. The physical parameters 
$M,\ldots,K_\mathrm{s}$ in \eqref{eq:unbalanced} are taken from \cite{Kulcsar2009}.

By applying RK4-based numerical integration with $T_\mathrm{s}$, \eqref{eq:unbalanced} is simulated for a ZOH input signal $v_t$ generated by a DT multisine $u_{k}$ with 16666 components in the range $[0, 5]$ Hz with a uniformly distributed phase in $ [0, 2\pi]$. The resulting $u$ is limited to an amplitude of 3 V to prevent the unbalanced disc from going over the top. The sampled output measurements $y_k$ are perturbed by $e_k$ resulting in a \emph{signal-to-noise ratio} (SNR) of 30 dB. This provides a data sequence $\mathcal{D}_N$ of size $N =10^5$. The data sequence is split into estimation, validation, and test data sets with size $N_{\text{est}} = 6\cdot10^4$, $N_{\text{val}} = 2\cdot10^4$, $N_{\text{test}} = 2\cdot10^4$, respectively.



An ANN-SS model is identified as in \cite{pmlr-v144-beintema21a} with the hyperparameters shown in Table \ref{tab:Hyperparam}. The resulting model has \emph{normalized root mean square} error of 3.2\% on the test data set, corresponding to an accurate DT model of \eqref{eq:unbalanced}. 

\begin{table}[t]
    \centering
    \caption{Hyperparameters for identifying the ANN-SS model.} \vspace{-2mm}
    \begin{tabular}{c|c|c|c|c|c|c}
\!hidden layers\!&\!nodes\!&\!$n_\mathrm{x}$\!&\!$n_\mathrm{u}$ $n_\mathrm{y}$\!&\!$n_a$ $n_b$\!&\!epochs\!&\!batch size\!\\
\hline
2 & 64 & 2 & 1 & 4 & 250 & 256\\
\end{tabular}
    \label{tab:Hyperparam} \vspace{-5mm}
\end{table}

For the MPC, we select $Q_i = \text{diag}(10^3,10)$
, $R_i = 1$, 
with $-\hat{y}_\text{min}=\hat{y}_\text{max}=1.2$, $-u_\text{min}=u_\text{max}=4$, and control horizon $N_\mathrm{p} = 10$. LPV conversion \eqref{eq:abcd_def} for calculation of the MPC is set up by using Simpson's rule for integration and \texttt{functorch}\cite{functorch2021} for differentiation with $d\lambda = 0.05$ as the integration step size. 
The QP problem in terms of minimization of \eqref{eq:CostNonConstrained} under \eqref{eq:const:final} is solved with \texttt{osqp}\cite{Stellato2020}. Convergence in Step 8 of Algorithm \ref{algo:LPV_MPC} is defined as $\| \breve U_k^{(j)} - \breve U_k^{(j-1)} 
\|_2 \leq 0.1$ with $\breve U_k^{(j)}$ denoting the optimal solution of \eqref{eq:CostNonConstrained} under \eqref{eq:const:final} at iteration $j$.

\begin{figure}
    \centering
    \includegraphics[width=0.45\textwidth]{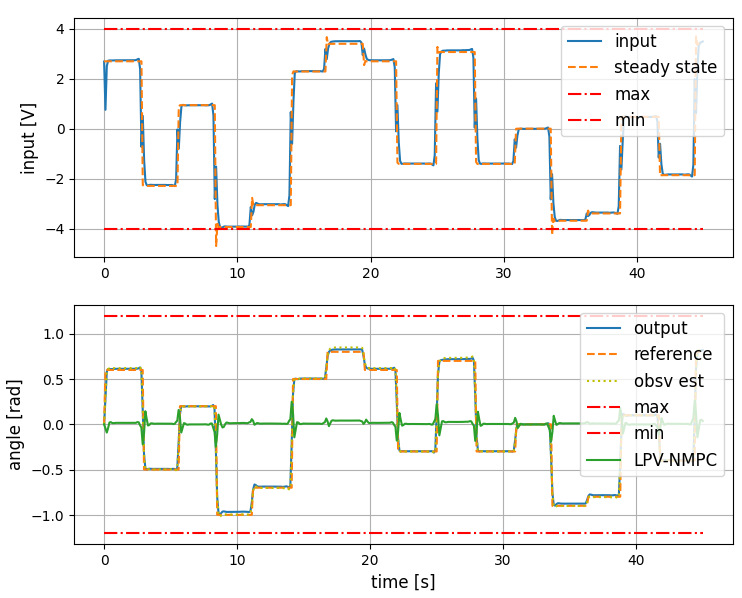} \vspace{-3mm}
    \caption{Results of the proposed MPC algorithm, based on the identified ANN-SS model, applied on \eqref{eq:unbalanced}. The plot shows the demanded piece-wise constant reference, noiseless output of \eqref{eq:unbalanced}, resulting input commands, the encoder estimated states based output of the ANN-SS model, the input and output constraints and the difference between the proposed MPC approach and the full NMPC solution of \eqref{eq:nl_ocp}.}
    \label{fig:StatesInputMPC.} \vspace{-4mm}
\end{figure}

Fig. \ref{fig:StatesInputMPC.} shows the resulting response of system \eqref{eq:unbalanced} controlled by the proposed  MPC algorithm using the identified ANN-SS model, while tracking a piece-wise constant reference signal for the disk angle. As we can see, the proposed MPC solution provides a reliable tracking of the reference and when compared to the  full NMPC solution of \eqref{eq:nl_ocp} using the \texttt{ipopt} solver \cite{ipopt2006}, the difference between the obtained responses is relatively small. We can also see that both MPC controllers are capable of tracking the reference as long as the input required is within the region on which the encoder and the ANN-SS model were trained. If the required input  falls outside that region a small steady-state error occurs. This can be explained by model errors in the networks for these inputs due to model extrapolation and generally the same behavior is expected from any identified black-box model on the considered data set.

While the proposed LPV-MPC method based solution and the NMPC produce relatively the same performance, the required computation times\footnote{Simulated on Intel(R) Core(TM) i7-7700HQ CPU @ 2.80GHz}, given in Table \ref{tab:computation_times}, show that the proposed QP iterations are much more efficient than the general \texttt{ipopt} solver-based NMPC solution. Furthermore, out of the 450 simulation steps, 338 required only a single iteration, 110 had two iterations, and only 2 required three steps to converge. The overall computation time is dominated by the LPV conversion, while the solver takes relatively little time. Furthermore, while there are large maximum values in the computation times, the average and standard deviation are relatively low, showing efficiency of the proposed scheme.

\begin{table}[t]
    \centering
    \caption{Computational time} \vspace{-2mm}
    \label{tab:computation_times}
    \begin{tabular}{l|r|r|r|r} 
        & Max time & Average time & St. dev. & Solver time \\
        \hline
        LPV-MPC & $65.0 \mathrm{~ms}$ & $14.1 \mathrm{~ms}$ \hspace{2mm} & $7.07 \mathrm{~ms}$ & $4.79 \mathrm{~ms}$\hspace{1mm} \\
        NMPC ipopt & $100.0 \mathrm{~ms}$ & $18.1 \mathrm{~ms}$  \hspace{2mm} & $5.75 \mathrm{~ms}$ & $16.30 \mathrm{~ms}$\hspace{1mm}  \\
\end{tabular} \vspace{-1mm}
\end{table}

\section{Conclusions}

For identified ANN-SS models, we have proposed an efficient iterative LPV-MPC solution which is based on two key ingredients: (i) approximation-free conversion of ANN-SS models to an LPV form through the Fundamental Theorem of Calculus, which can be efficiently computed via autograd methods and multi-threaded numerical integration, (ii) direct formulation of successive QP problems using the obtained optimal solution from the previous iteration step as the scheduling for the LPV prediction model. We have shown that computational load of the proposed MPC solution is drastically lower than a general NMPC method applied with the ANN-SS model thanks to the approximation-free LPV form compared to local linearizations of the problem and the computational tool-chain that heavily relies on the structure of the ANNs and the available autograd tools. Furthermore, the approach  benefits from the encoder structure provided by recent ANN-SS estimation methods like SUBNET, which enables reliable initial state calculation for the MPC scheme based on noisy measurements from the system.

\bibliographystyle{IEEEtran}
\bibliography{refs}

\begin{thebibliography}{10}
\providecommand{\url}[1]{#1}
\csname url@rmstyle\endcsname
\providecommand{\newblock}{\relax}
\providecommand{\bibinfo}[2]{#2}
\providecommand\BIBentrySTDinterwordspacing{\spaceskip=0pt\relax}
\providecommand\BIBentryALTinterwordstretchfactor{4}
\providecommand\BIBentryALTinterwordspacing{\spaceskip=\fontdimen2\font plus
\BIBentryALTinterwordstretchfactor\fontdimen3\font minus
  \fontdimen4\font\relax}
\providecommand\BIBforeignlanguage[2]{{%
\expandafter\ifx\csname l@#1\endcsname\relax
\typeout{** WARNING: IEEEtran.bst: No hyphenation pattern has been}%
\typeout{** loaded for the language `#1'. Using the pattern for}%
\typeout{** the default language instead.}%
\else
\language=\csname l@#1\endcsname
\fi
#2}}

\bibitem{rawlings_mpc}
J.~B. Rawlings, D.~Q. Mayne, and M.~M. Diehl, \emph{Model Predictive Control:
  Theory, Computation, and Design}.\hskip 1em plus 0.5em minus 0.4em\relax Nob
  Hill Pub., 2017.

\bibitem{Verschueren2022}
R.~Verschueren, G.~Frison, D.~Kouzoupis, J.~Frey, N.~v. Duijkeren, A.~Zanelli,
  B.~Novoselnik, T.~Albin, R.~Quirynen, and M.~Diehl, ``{Acados—a Modular
  Open-Source Framework for Fast Embedded Optimal Control},'' \emph{Math. Prog.
  Comp.}, vol.~14, no.~1, pp. 147--183, 2022.

\bibitem{Limon2008}
D.~Limon, A.~Ferramosca, I.~Alvarado, and T.~Alamo, ``Nonlinear {MPC} for
  tracking piece-wise constant reference signals,'' \emph{IEEE Trans. Aut.
  Cont.}, vol.~63, p. 3735–3750, 2018.

\bibitem{suykens1995}
J.~A.~K. Suykens, B.~L.~R. De~Moor, and J.~Vandewalle, ``Nonlinear system
  identification using neural state space models, applicable to robust control
  design,'' \emph{Int. J. of Control}, vol.~62, no.~1, pp. 129--152, 1995.

\bibitem{Beintema2022}
G.~I. Beintema, M.~Schoukens, and R.~Tóth, ``Deep subspace encoders for
  nonlinear system identification,'' arXiv: 2210.14816, 2022.

\bibitem{Forgione2021TruncSimulation}
M.~Forgione and D.~Piga, ``Continuous-time system identification with neural
  networks: Model structures and fitting criteria,'' \emph{Eu. J. of Cont.},
  vol.~59, pp. 69--81, 2021.

\bibitem{lazar2002}
M.~Lazar and O.~Pastravanu, ``A neural predictive controller for non-linear
  systems,'' \emph{Math. and Comp. in Sim,}, vol.~60, pp. 315--324, 2002.

\bibitem{Lawrynczuk2014}
M.~Lawrynczuk, \emph{Computationally Efficient Model Predictive Control
  Algorithms: A Neural Network Approach}.\hskip 1em plus 0.5em minus
  0.4em\relax Springer, 2014.

\bibitem{ipopt2006}
A.~W{\"a}chter and L.~Biegler, ``On the implementation of an
  interior\nobreakdash-filter line\nobreakdash-search algorithm for
  large\nobreakdash-scale nonlinear programming,'' \emph{Math. Prog.}, vol.
  106, pp. 25--56, 2006.

\bibitem{masti2021autoencoders}
D.~Masti and A.~Bemporad, ``Learning nonlinear state--space models using
  autoencoders,'' \emph{Automatica}, vol. 129, p. 109666, 2021.

\bibitem{GonzalezCisneros2021}
P.~Cisneros, ``Quasi-linear model predictive control: Stability, modelling and
  implementation,'' Ph.D. dissertation, Technical University Hamburg, 2021.

\bibitem{hashemi2012}
S.~M. Hashemi, H.~S. Abbas, and H.~Werner, ``Low-complexity linear
  parameter-varying modeling and control of a robotic manipulator,'' \emph{Con.
  Eng. Pract.}, 2012.

\bibitem{Marcos04}
A.~Marcos and G.~J. Balas, ``Development of linear-parameter-varying models for
  aircraft,'' \emph{J. of Guid., Cont. and Dyn.}, vol.~27, no.~2, pp. 218--228,
  2004.

\bibitem{Kwiatkowski4494453}
A.~Kwiatkowski and H.~Werner, ``{PCA}-based parameter set mappings for {LPV}
  models with fewer parameters and less overbounding,'' \emph{IEEE Trans. on
  Cont. Sys. Tech.}, vol.~16, no.~4, pp. 781--788, 2008.

\bibitem{Toth20bIET}
A.~Sadeghzadeh, B.~Sharif, and R.~T\'{o}th, ``Affine linear parameter-varying
  embedding of non-linear models with improved accuracy and minimal
  overbounding,'' \emph{IET Cont. Theory \& App.}, vol.~14, pp. 3363--3373,
  2020.

\bibitem{Koelewijn2023}
P.~J. Koelewijn, ``Analysis and control of nonlinear systems with stability and
  performance guarantees: A linear parameter-varying approach,'' Ph.D.
  dissertation, Eindhoven University of Technology, 2023.

\bibitem{shalev2014Subgradient}
S.~Shalev-Shwartz and S.~Ben-David, \emph{Understanding machine learning: From
  theory to algorithms}.\hskip 1em plus 0.5em minus 0.4em\relax Cambridge
  University Press, 2014.

\bibitem{Hespe2021}
C.~Hespe and H.~Werner, ``Convergence properties of fast quasi-{LPV} model
  predictive control,'' in \emph{Proc. of the 60th IEEE Conf. on Decision and
  Cont.}, 2021, pp. 3869--3874.

\bibitem{boggs_tolle_1995}
P.~T. Boggs and J.~W. Tolle, ``Sequential quadratic programming,'' \emph{Acta
  Numerica}, vol.~4, p. 1–51, 1995.

\bibitem{Kulcsar2009}
B.~Kulcs\'{a}r, J.~Dong, J.~W. {van Wingerden}, and M.~Verhaegen, ``{LPV}
  subspace identification of a {DC} motor with unbalanced disc,'' in
  \emph{Proc. of the 15th IFAC Symp. on Sys. Id.}, 2009, pp. 856--861.

\bibitem{pmlr-v144-beintema21a}
G.~Beintema, R.~Toth, and M.~Schoukens, ``Nonlinear state-space identification
  using deep encoder networks,'' in \emph{Proc. of the 3rd Conf. on Learn. for
  Dyn. and Cont.}, ser. Proceedings of Machine Learning Research, vol. 144,
  2021, pp. 241--250.

\bibitem{functorch2021}
R.~Z. Horace~He, ``functorch: Jax-like composable function transforms for
  pytorch,'' \url{https://github.com/pytorch/functorch}, 2021.

\bibitem{Stellato2020}
B.~Stellato, G.~Banjac, P.~Goulart, A.~Bemporad, and S.~Boyd, ``{OSQP}: an
  operator splitting solver for quadratic programs,'' \emph{Math. Prog. Comp.},
  vol.~12, no.~4, pp. 637--672, 2020.

\end{thebibliography}

\end{document}